\documentclass[12pt]{paper}
\usepackage[utf8]{inputenc}
\usepackage{amsmath}
\usepackage{amssymb,amsfonts}
\usepackage[T1]{fontenc}
\usepackage[english]{babel}
\usepackage{color}
\usepackage{array}
\usepackage{hhline}
\usepackage{hyperref}
\usepackage{csquotes}
\usepackage{fancyhdr}
\usepackage[square,numbers]{natbib}
\hypersetup{colorlinks=true, linkcolor=blue, citecolor=blue, filecolor=blue, urlcolor=blue}
\usepackage{graphicx}

\makeatother
\date{}
\begin{document}
	\title{\bfseries No Eternal Collapse in General Relativity}
	\author{Chandra Prakash}
	\maketitle
	\thispagestyle{empty}

\begin{abstract}
A critical overview of the concepts that govern the principles of the Eternally Collapsing Object (ECO) is entailed in this article. The mathematical analysis will be dealt to conclude the proof required for the ECO paradigm is ad-hoc in nature at best. The first section details in rejecting black holes which considers the possibility of formation of trapped surfaces,  so we will begin our work by looking into \textquote{non occurrence of trapped surfaces}. After this a detailed analysis on the claim that the mass of the Schwarzschild black hole being zero is in actuality a valid claim or not. The main aim of this article is to address the core issue of whether black holes indeed do exist or is it a consequence of biases that circumvent common sense over the years. This article aims to bridge the gap between ECOs and black holes and if at all black holes do exist, which was vehemently denied by the person whose paper I am reviewing. The analysis presented here will show us, why ECOs are baseless and can not really be considered the solution to black hole problem.
\end{abstract}
	
\section*{Introduction}
Eternally Collapsing Object or ECOs for short, is an alternative solution to Black Hole Problem which was proposed by Dr. Abhas Mitra. ECOs are interesting objects because they can potentially mimic a black hole and even explain effects such as hawking radiation. They are so dense that their collision could explain what was observed during the gravitational wave detection at LIGO. Existence of ECO and the non existence of black hole can indeed solve a lot of issue and give us all a sense of relief that now we understand a lot more about the universe and there are few less mystery to unravel! But before we consider ECOs to be real and black holes as fairy tale we need to look into trapped surfaces, their occurrence or non-occurrence, Black Hole Mass Problem and the General Relativistic Star which all together form the basis of ECO. ECOs are nothing but ever collapsing star, the proper time taken by such a star to complete its collapse is infinite and by the time it reaches that condition, star has already had radiated everything it had. We review here the mathematical analysis which lead to this conclusion and discover for ourselves if such a solution is actually viable and based on undeniable mathematical rigour or birth of misconception, meanwhile we will be using natural system where $G = c = 1$ unless otherwise stated. First section of this paper deals with the non occurrence of trapped surfaces, where we present few different ways which leads to the incompatibility of event horizon with \textquote{Equivalence Principle}. However we soon find all such approaches have few steps in error and can be eliminated from possible inconsistencies. Section 2 of this paper deals with Black Hole Mass problem, where we explore few different ways of how schwartzschild black holes are actually massless. Section 3 then explores few ways of defining ECOs and in section 4 we conclude the whole paper discussing how ECOs are not really a solution to any problem but merely an inquisitive doubt.

\section{Non-Occurrence of Trapped Surfaces}
In 1965, Sir Roger Penrose was exploring the possibility of black hole formation where the spherical symmetry was not inherently assumed, exploring the possible ways and condition for the singularity to form, he defined the notion of closed trapped surfaces as the exact condition when any deviation from spherical symmetry will not be able to stop the  gravitational collapse.\cite{PhysRevLett.14.57} So it seems reasonable to start our inquiry with the formation of trapped surfaces and see how it is inconsistent general relativity. To begin we needed to define few quantities and use the constraint defined on them\cite{may1966hydrodynamic} to arrive at some conclusion 

\begin{equation}
\Gamma^2 = 1 + U^2 - \frac{2GM}{R}
\end{equation}

The above equation could be exploited to prove that a black hole never forms, if somehow we managed to arrive at expression like $R \ge 2GM$. Which is possible in few ways if we define our $\Gamma$ and $U$ to be something like this\cite{mitra2000non}:

\begin{equation}
\Gamma = \frac{dR}{dl} \quad\text{and}\quad U = \frac{dR}{d\tau} = \frac{dR}{dl}\frac{dl}{d\tau}
\end{equation}

However the error in reasoning lies at being unable to recognize that both $\Gamma$ and $U$ are originally defined in terms of $\partial$ derivatives not total derivative, but still we can explore this if somehow we manage to justify the reasoning. The reasoning needed to justify the adhoc result was soon found and presented as a special case where the calculation was concerned with a particular trajectory being defined by the expression $dR = 0$\cite{mitra2006black}. The result now does seem to make sense but it does not imply anything we already didn't know. Both the $\Gamma$ and $U$ were originally defined in \textquote{co-moving coordinate system}\cite{may1966hydrodynamic}.

\begin{equation}
dR = 0 \implies \frac{dr}{dt} = -\frac{\dot{R}}{R'} \label{dR=0}
\end{equation}

Regardless of how we use expression (\ref{dR=0}), to arrive at $R \ge 2GM$, it still doesn't prove the \textquote{non-occurrence of trapped surfaces}, since \textbf{$R(l,\tau)$} is defined in comoving coordinates\cite{may1966hydrodynamic}. If we try to express the equation in terms of $r$ (defined in non comoving coordinates) we arrive at results where $\dot{r} = 0$, which means the collapse never begun in the first place or the observer was standing outside the event horizon.

The other half attempts at same seems to fail because of misunderstanding the notations. In 1939, when Oppenheimer and Snyder first published their work on gravitational collapse, they arrived at few expressions describing the collapse\cite{oppenheimer1939continued} which are:
\begin{align*}
t(r, \tau) = \frac{2}{3} \sqrt{R}_{_{gb}} \left(\sqrt{r_{b}^{3}} - \sqrt{R_{gb}^{3}y^{3}}\right)
- 2R_{_{gb}}\sqrt{y} + R_{_{gb}}\ln\frac{\sqrt{y} + 1}{\sqrt{y} - 1} \\
\\
\text{where,} \quad
y = \frac{1}{2} \left[\left(\frac{r}{r_{b}}\right)^2 - 1 \right] + \frac{r_{b}R}{R_{_{gb}}r} \qquad\text{and} \qquad R_{_{gb}}=2GM
\end{align*}

\noindent From the expression it's obvious that $y>1$ and: \cite{mitra2006black} 
\begin{equation} \label{eq:1}
y_{r=r_{b}} =  \frac{R}{R_{_{gb}}}  > 1
\end{equation}

But in no way this expression says that trapped surface will never form, far from that, to actually interpret this expression we would need to look at equation (27) of \cite{oppenheimer1939continued}, which makes the equation (\ref{eq:1})

$$\left( \frac{-3 \sqrt{2GM}} {2}\tau +
\sqrt{r^3}\right)^{2/3} > 2GM$$

\subsection{Event Horizon is true singularity!}
The null hypersurface with $z=\infty$ defined at $R=2GM$ in schwartzschild metric is actually a coordinate singularity because it can be removed via coordinate transformation but does that make it physically non-singular? If we look at the expression for speed of any particle and take a limit as $R\rightarrow 2GM$ (even in Eddington Finkelstein coordinate), what we find is that all observers seem to move at speed of light at this point which is in direct contradiction with principle of relativity\cite{mitra2006black}. 

\begin{equation}
	\upsilon^2 = \frac{\left(g_{_{01}}^2 - g_{_{11}}g_{_{00}} \right)	\left(\frac{dx^1}{dx^0}\right)^2}{\left( g_{_{00}} + g_{_{01}}\frac{dx^1}{dx^0}\right)^2} \implies \lim_{{R}\rightarrow2GM} v^2 = 1
\end{equation}

Is there any way to reconcile with this issue? Infact there is, the above expression used in the calculation was originally defined by Landau and Lifshitz\cite{landau2013classical} with an external assumption that the light we used for synchronizing clocks experienced no redshift! However that rules out the possibility of using this expression where redshift can no longer be ignored. Infinite redshift at event horizon makes it impossible for us to use this equation to arrive at any sensible result.

Similar approach can also be seen at other places where authors exploit the equation expressing the acceleration due to gravity which diverges at event horizon\cite{mitra2006black}, but the expression used again to show this\cite{doughty1981acceleration} had already assumed a static timelike observer but since no observer can stay at rest on event horizon, this eliminates the possibility of using it for any further analysis at $R=2GM$ and thereby debunking all the claims regarding why \textquote{event horizon can not be real.}

\begin{equation}
a^R = \frac{GM}{R^2}\left(1 - \frac{2GM}{R}\right)^{-1/2}
\end{equation}

There is however one point of concern with event horizon that we can use invariants such as Karlhede Invariant to detect event horizon, but that can be resolved by looking at the same invarient in kerr-newmann spacetime which takes the form of\cite{moffat2014karlhede}:

\begin{equation}
\mathcal{I} = \frac{-720M^2(a^2cos^2\theta + R^2 - 2MR)Q_{1}Q_{2}}{(a^2cos^2\theta + R^2)^9}
\end{equation}
with
\begin{equation*}
Q_{1} = (acos\theta -R)^4 - 4aR^2 cos\theta(3acos\theta - 2R),
\end{equation*},
\begin{equation*}
Q_{2} = (acos\theta -R)^4 - 4a^2R cos^2\theta(3R - 2acos\theta).
\end{equation*}

Expressions such as above opens up the possibility that event horizon is no longer just the hypersurface with infinite redshift but a membrane of somekind with interesting quantum effects. This is a whole new field of investigation being explored by several physicists\cite{price1988membrane,thorne1986membrane}. There may indeed be several invariants which can be used to detect event horizon but none of them rule out event horizon as being inconsistent with General Relativity but add additional structure to the event horizon such as \textquote{firewall proposal}\cite{moffat2014karlhede}.

\section{Massless Black Hole}
The interpretation of black hole mass had been a troublesome issue for some time, we have had expressions like komar integral or bondi mass and even ADM mass but all of them are defined for asymptotic observers. What does the $M$ term in schwartzschild metric actually represent and what does the mass of black hole even mean? Are the claims regarding the $M$ term for black hole to be actually zero, true in some regard?
The answer to all these problem is no, the only black hole with zero \textquote{mass} is point particle\cite{arnowitt1960finite}. When we have multiparticle system we can no longer use the same expression to calculate the self energy and hence claiming that the same result holds true even for many particle system is incautious. However there are several approaches to show that this result holds true in general, all of them wrong. We will review few of them one by one:

\subsection{Oppenheimer-Tolman-Volkoff Equation}
1939 paper of Oppenheimer and Snyder concerning the gravitational collapse had already assumed the $p=0$ to hold throughout the dust, however applying this result in Oppenheimer-Tolman-Volkoff Equation\cite{oppenheimer1939massive} we soon arrive at $\rho=0$\cite{mitra2006black}.
\begin{equation}\label{op}
\frac{dp}{dR} = - G \frac{M(r) + 4\pi p R^3 }{R^2 \left( 1 - \frac{2GM}{R^2}\right)} [p + \rho]
\end{equation} 

However the equation (\ref{op}) holds true only for hydrostatic equilibrium which won't be applicable for collapsing fluid.

\subsection{Invariance of 4-volume}
General Relativity is written in the language of tensors which makes it coordinate independent. There are several quantities which do not depend on coordinate system in general relativity, 4-volume is one of them. Comparing this invariant volume integral we could show that the mass of black hole is zero\cite{mitra2006black}. However the analysis seems to have missed the point that invarient volume integral is expressed in terms of differential forms on differential elements from ordinary calculus\cite{carroll2019spacetime}. Repeating the same calculation in proper way we arrive at\cite{doi:10.1063/1.5011133}:

\begin{equation*}
\int R^2~ dT \wedge dR + \int R^2 \frac{\alpha_{0}}{R - \alpha_{0}}dR \wedge dR = \int R^2~dT\wedge dR
\end{equation*}

\bigskip

\noindent It is a well known result in differential geometry that $  dR \wedge dR = 0$, leading to

\begin{equation*}
\int R^2~ dT \wedge dR = \int R^2~dT\wedge dR
\end{equation*}
The result is, as what one would expect. There is nothing much to draw from this..

\subsection{Mass of Oppenheimer Snyder Black Hole}
Oppenheimer and Snyder did their work on gravitational collapse in comoving coordinate system and used birchoff theorem in the end to relate the parameters to non comoving coordinate. It is only reasonable we do the same, but for mass integral!
\begin{equation}
 \text{ Mass = }\int T^{00}d\mathcal{V} = \int \rho d\mathcal{V} = \int \frac{\rho}{1 - \nu^2}d\mathcal{V'}
\end{equation}
However if we use this equation to compare the results in comoving and non-comoving coordinate system, we arrive at:
\begin{equation*}
\int_{0}^{R_{b}} 4 \pi \left(  \rho(r,t) - \frac{\rho(R,T)  }{1 - v^2} \right) R^2 dR = 0
\end{equation*}
Since $\rho$ is a scalar, we expect it to be same in both coordinate system and thus arrive at $\rho \nu^2 =0$ which leads $\rho=0$\cite{mitra2011fallacy} but before jumping to conclusion let us do another calculation, just to make sure we are interpreting the results in correct way. We can just as well compare  
$$P^{\mu}P_{\mu} = m^{2}$$
Evaluating it in a coordinate system where the particle has non zero speed and in another system where it is at rest
\begin{equation}
E^{2} -\vec{P}\cdot\vec{P} =m^{2} \qquad \& \qquad E^{2} =m^{2}
\end{equation}
But since the particle mass is same and Energy is a scalar it would remain invarient, upon substituting we will arrive at the bizzare conclusion that the momentum of particle in the frame where it had non zero speed is infact "0".
$$E^{2} -\vec{P}\cdot\vec{P} = E^2$$

\noindent The easiest way to recognize the error is to use different notation for both $E$'s since they are both being expressed in different coordinate system (frame of reference) and hence should have been treated that way even though they look like scalars, they are not! They are components of rank 1 tensor. This leads to the natural conclusion that it is $\nu=0$ not $\rho$ and thereby proving we had a non-collapsing fluid!

\section{Entering the ECO paradigm}
If we had already established that the true mass of Black Holes are zero and trapped surfaces never form. It would have been a natural conclusion that the collapse of any general relativistic star never reaches the $R=2GM$ point. Since it is not the case, we are only going to look into the general relativistic star as an ECO model\cite{mitra2006radiation}. The analysis begins by first proposing the form of Eddington Limit which is supposed to hold even in general relativity. This expression comes from the constraint defined as \cite{mitra1998maximum}:

\begin{equation}\label{eq11}
L_{\infty} = \frac{L_{0}}{(1+z)^2} \implies L_{0} = \frac{4\pi GMm_{0}c}{\sigma}(1+z)
\end{equation}
If we use the form of $L_{0}$ defined in above equation, we arrive at the general result defined in \cite{mitra2006radiation} as:

\begin{equation}
\frac{p_{r}}{p_{0}} = \frac{m_{0}c^2}{3kT} \frac{\rho_{r}}{ \rho_{0} } = \frac{m_{0}c^2}{3kT} \frac{\alpha GM }{2Rc^2} (1 + z)
\end{equation}
This form suggests that as $z\rightarrow \infty$ pressure tends to blow up, thereby making the general relativistic star possible and ruling out all other possibilities which involves assuming vanishing pressure in spherical collapse! There is just one ambiguity with the result, it doesn't work if we take the value of $L_{0}$ aa this:
\begin{equation*}
L_{\infty} = \frac{L}{(1 + z)^2}\\
\implies
L_{0} = \frac{4\pi GM m_{0} c}{\sigma }
\end{equation*}
This too satisfies the boundary condition defined in equation (\ref{eq11}). There are several ways to arrive at this result but all of them become inconsistent with the ECO paradigm in one way or the another. Leiter and Robertson\cite{leiter2003does} tried to do the same, they made few mistakes in the equation A9, misinterpreted the result from \cite{lindquist1966relativistic} or \cite{hernandez1966observer}. 	

\subsection[Supermassive Stars and Eddington Luminosity]{\centering Supermassive Stars and Eddington Luminosity}
Supermassive Stars are interesting problem in astrophysics, not because of it's mass but mainly because it involves general relativity to prove it's stability. Weinberg's book on Gravitation and Cosmology also has a section devoted to this, where he explicitly mentions that the supermassive stars are newtonian star! Here what concerns our problem is the question of eddington luminosity and that is what we are gonna focus on! We begin our analysis by first assuming a star with zero gas pressure and is completely being supported because of the radiation pressure. 
\begin{align*}
-\frac{\nabla p}{\rho} - \nabla \phi = 0 \implies
\frac{\nabla p}{\rho} = - \nabla \phi
\end{align*}
If we want to translate the same derivation to general relativity then we will need to use relativistic fluid equation for hydrostatic equilibrium\cite{landau2013fluid}

\begin{equation}
\frac{1}{\rho c^2} \frac{\partial p}{\partial x^\alpha} = -\frac{1}{2} \frac{\partial }{\partial x^\alpha}\log(g_{_{00}})
\end{equation}

\bigskip
\noindent We can use this expression to arrive at Eddington Luminosity at the boundary of the star where interior solution matches the exterior solution due to birchoff theorem, even in General Relativistic scenario, since $\frac{1}{\rho} \frac{\partial p}{\partial x^\alpha} = -\frac{\kappa}{c}F_{rad}=-\frac{\kappa}{c}\frac{\partial^2 E}{\partial A \partial t}$\cite{rybicki1986radiative}:
\begin{align}\label{14}
L_{ed}(\infty) = \frac{1}{(1+z)^2}\int F_{rad}\cdot dS &= \frac{1}{(1+z)^2}\int \frac{c^3}{2\kappa}\frac{\partial }{\partial x^\alpha}\log(g_{_{00}}) \cdot dS \notag\\
\notag\\
&=\frac{1}{(1+z)^2}\frac{c}{\kappa}\frac{4\pi GM}{\left(1 - \frac{2GM}{rc^2}\right)} = \frac{1}{(1+z)^2}\frac{4\pi GMc}{\kappa}(1+z)^2
\end{align}
In the last step we used the redshift factor $1+z= \left(1-\frac{2GM}{r}\right)^{-1/2}$\cite{weinberg1972gravitation}, but it doesn't quite match with the boundary condition of the equation (\ref{eq11}). This is my point of concern, taking a look at the expression, part of it does not only
blow up as $z \rightarrow \infty$ but another part of it vanishes as well making it indeterminate in nature. In the limiting case  $z\rightarrow \infty$ of Equation \ref{14}, we find that the stellar remnant has finite nonzero luminosity of $\frac{4\pi GMc}{\kappa}$, in direct contradiction with Black Hole Paradigm. In fact there are studies which suggests the "General Relativistic Instability of Supermassive Stars" and can be used together with equation (\ref{14}) to make the concept of Eddington Luminosity a rather non-relativistic element.

\section{Conclusion}
In the first section of the paper, we found out that the arguments regarding the non occurrence of trapped surfaces were based on ignoring the fact that all three parameters involved in the analysis were originally defined in comoving coordinate system. In second section we saw the conditions for black holes to be massless were because of calculation error and in third section we explored the general relativistic star only to conclude that the result is based on personal bais and completely circumstantial expression. 

Non-Existence of Trapped Surface could have opened up a door for eternally collapsing radiating star which attains the condition for $M=0$ via radiation. But the possibility of losing all of stellar mass due to radiation is in violation with conservation of quantum numbers. If a radiating baryonic star has mass $M_{b}$ due to existence of baryons, then the metric describing a radiating star like Vaidya metric\cite{vaidya1951gravitational} will take the same form as Schwarzschild or Eddington-Finkelstein metric when the star looses all of kinetic energy in the form of radiation attaining the case of $E = M_{b}c^2$. At this point no more photons can be emitted and hence no massloss. This will convert the vaidya metric back to Eddington-Finkelstein metric and with that all the work concerning Black Hole will come back into the picture.

However before trapped surface really forms, there is a possibility that the star could somehow attain equilibrium but the analysis showing the possibility for such an event to occur seems to be using an equation which is purely coincidental in that it just meets the boundary condition and there is no clear reason for us to believe in such equation. Using other equation that meets the conditions never leads to the same conclusion. All of proof is based on somehow getting the $z$ factor in numerator and using non-relativistic thermodynamics to argue the existence of diverging radiation and then justifying it with ad-hoc calculation. In non-relativistic weak gravitation case, gas pressure dominates the overall radiation pressure and keeps the star in equilibrium but that can't happen in strong gravitational case. The radiation pressure may seem to dominate but it doesn't actually counteract the gravity just like degeneracy pressure. 

The above discussion makes ECO baseless and their existence very much unlikely as there are no clear mathematical analysis showing why such stars should even exist in the first place. The work of eternally collapsing object could have one place which doesn't voilate any conservation laws and that is a star made entirely of photons or massless particles.
\bibliographystyle{abbrvnat}
\bibliography{citation}

\begin{thebibliography}{24}
\providecommand{\natexlab}[1]{#1}
\providecommand{\url}[1]{\texttt{#1}}
\expandafter\ifx\csname urlstyle\endcsname\relax
  \providecommand{\doi}[1]{doi: #1}\else
  \providecommand{\doi}{doi: \begingroup \urlstyle{rm}\Url}\fi

\bibitem[Arnowitt et~al.(1960)Arnowitt, Deser, and Misner]{arnowitt1960finite}
R.~Arnowitt, S.~Deser, and C.~Misner.
\newblock Finite self-energy of classical point particles.
\newblock \emph{Physical Review Letters}, 4\penalty0 (7):\penalty0 375, 1960.

\bibitem[Carroll(2019)]{carroll2019spacetime}
S.~M. Carroll.
\newblock \emph{Spacetime and geometry}.
\newblock Cambridge University Press, 2019.

\bibitem[Doughty(1981)]{doughty1981acceleration}
N.~A. Doughty.
\newblock Acceleration of a static observer near the event horizon of a static
  isolated black hole.
\newblock \emph{American Journal of Physics}, 49\penalty0 (5):\penalty0
  412--416, 1981.

\bibitem[Hernandez~Jr and Misner(1966)]{hernandez1966observer}
W.~C. Hernandez~Jr and C.~W. Misner.
\newblock Observer time as a coordinate in relativistic spherical
  hydrodynamics.
\newblock \emph{The Astrophysical Journal}, 143:\penalty0 452, 1966.

\bibitem[Kundu(2017)]{doi:10.1063/1.5011133}
P.~K. Kundu.
\newblock Comment on “comments on ‘the euclidean gravitational action as
  black hole entropy, singularities and space-time voids’” [j. math. phys.
  50, 042502 (2009)]–schwarzschild black hole lives to fight another day.
\newblock \emph{Journal of Mathematical Physics}, 58\penalty0 (11):\penalty0
  114101, 2017.
\newblock \doi{10.1063/1.5011133}.
\newblock URL \url{https://doi.org/10.1063/1.5011133}.

\bibitem[Landau and Lifshitz(2013)]{landau2013fluid}
L.~Landau and E.~Lifshitz.
\newblock \emph{Fluid Mechanics: Landau and Lifshitz: Course of Theoretical
  Physics, Volume 6}.
\newblock Number v. 6. Elsevier Science, 2013.
\newblock ISBN 9781483161044.
\newblock URL \url{https://books.google.co.in/books?id=eOBbAwAAQBAJ}.

\bibitem[Landau(2013)]{landau2013classical}
L.~D. Landau.
\newblock \emph{The classical theory of fields}, volume~2.
\newblock Elsevier, 2013.

\bibitem[Leiter and Robertson(2003)]{leiter2003does}
D.~Leiter and S.~Robertson.
\newblock Does the principle of equivalence prohibit trapped surfaces from
  forming in the general relativistic collapse process?
\newblock \emph{Foundations of Physics Letters}, 16\penalty0 (2):\penalty0
  143--161, 2003.

\bibitem[Lindquist(1966)]{lindquist1966relativistic}
R.~W. Lindquist.
\newblock Relativistic transport theory.
\newblock \emph{Annals of physics}, 37\penalty0 (3):\penalty0 487--518, 1966.

\bibitem[May and White(1966)]{may1966hydrodynamic}
M.~M. May and R.~H. White.
\newblock Hydrodynamic calculations of general-relativistic collapse.
\newblock \emph{Physical Review}, 141\penalty0 (4):\penalty0 1232, 1966.

\bibitem[Mitra(1998)]{mitra1998maximum}
A.~Mitra.
\newblock Maximum accretion efficiency in general theory of relativity.
\newblock \emph{arXiv preprint astro-ph/9811402}, 1998.

\bibitem[Mitra(2000)]{mitra2000non}
A.~Mitra.
\newblock Non-occurrence of trapped surfaces and black holes in spherical
  gravitational collapse.
\newblock \emph{Foundations of Physics Letters}, 13\penalty0 (6):\penalty0
  543--579, 2000.

\bibitem[Mitra(2006{\natexlab{a}})]{mitra2006black}
A.~Mitra.
\newblock Black holes or eternally collapsing objects: a review of 90 years of
  misconceptions.
\newblock \emph{Focus on Black Hole Research,(ed. PV Kreitler, Nova Sc.
  Publishers, NY)}, 2006{\natexlab{a}}.

\bibitem[Mitra(2006{\natexlab{b}})]{mitra2006radiation}
A.~Mitra.
\newblock Radiation pressure supported stars in einstein gravity: eternally
  collapsing objects.
\newblock \emph{Monthly Notices of the Royal Astronomical Society},
  369\penalty0 (1):\penalty0 492--496, 2006{\natexlab{b}}.

\bibitem[Mitra(2011)]{mitra2011fallacy}
A.~Mitra.
\newblock The fallacy of oppenheimer snyder collapse: no general relativistic
  collapse at all, no black hole, no physical singularity.
\newblock \emph{Astrophysics and Space Science}, 332\penalty0 (1):\penalty0
  43--48, 2011.

\bibitem[Moffat and Toth(2014)]{moffat2014karlhede}
J.~Moffat and V.~Toth.
\newblock Karlhede's invariant and the black hole firewall proposal.
\newblock \emph{arXiv preprint arXiv:1404.1845}, 2014.

\bibitem[Oppenheimer and Snyder(1939)]{oppenheimer1939continued}
J.~R. Oppenheimer and H.~Snyder.
\newblock On continued gravitational contraction.
\newblock \emph{Physical Review}, 56\penalty0 (5):\penalty0 455, 1939.

\bibitem[Oppenheimer and Volkoff(1939)]{oppenheimer1939massive}
J.~R. Oppenheimer and G.~M. Volkoff.
\newblock On massive neutron cores.
\newblock \emph{Physical Review}, 55\penalty0 (4):\penalty0 374, 1939.

\bibitem[Penrose(1965)]{PhysRevLett.14.57}
R.~Penrose.
\newblock Gravitational collapse and space-time singularities.
\newblock \emph{Phys. Rev. Lett.}, 14:\penalty0 57--59, Jan 1965.
\newblock \doi{10.1103/PhysRevLett.14.57}.
\newblock URL \url{https://link.aps.org/doi/10.1103/PhysRevLett.14.57}.

\bibitem[Price and Thorne(1988)]{price1988membrane}
R.~H. Price and K.~S. Thorne.
\newblock The membrane paradigm for black holes.
\newblock \emph{Scientific American}, 258\penalty0 (4):\penalty0 69--77, 1988.

\bibitem[Rybicki and Lightman(1986)]{rybicki1986radiative}
G.~Rybicki and A.~Lightman.
\newblock Radiative processes in astrophysics, 1986.

\bibitem[Thorne et~al.(1986)Thorne, Price, and Macdonald]{thorne1986membrane}
K.~S. Thorne, R.~H. Price, and D.~A. Macdonald.
\newblock The membrane paradigm.
\newblock \emph{Yale University, New Haven, CT}, 6, 1986.

\bibitem[Vaidya(1951)]{vaidya1951gravitational}
P.~C. Vaidya.
\newblock The gravitational field of a radiating star.
\newblock In \emph{Proceedings of the Indian Academy of Sciences-Section A},
  volume~33, page 264. Springer, 1951.

\bibitem[Weinberg(1972)]{weinberg1972gravitation}
S.~Weinberg.
\newblock Gravitation and cosmology: principles and applications of the general
  theory of relativity.
\newblock 1972.

\end{thebibliography}
	
\end{document}